\documentclass[%
reprint,
amsmath,amssymb,
aps,
nofootinbib,
]{revtex4-2}

\usepackage{graphicx}
\usepackage{dcolumn}
\usepackage{bm}
\usepackage[percent]{overpic}
\usepackage{caption}
\usepackage{subcaption}
\usepackage{float}
\usepackage{xcolor}
\usepackage[normalem]{ulem}

\begin{document}

\preprint{APS/123-QED}

\title{Simple Magneto-Optical and Magnetic Traps for Dysprosium}


\author{Liam Domett-Potts$^1,^2$}

\author{Lucile Sanchez$^1,^2$}
\author{Charlotte Hayton$^1,^2$}
\author{Oscar Stone$^1,^2$}
\author{Nuttida Kaewart$^3$}
\author{Piyawat Chatchaichompu$^3$}
\author{Narupon Chattrapiban$^3$}
\author{Nithiwadee Thaicharoen$^3$}
\author{Mikkel  F. Andersen$^1,^2$}%
\affiliation{%
$^1$Department of Physics, University of Otago, Dunedin, New Zealand\\
$^2$The Dodd-Walls Centre for Photonic and Quantum Technologies, University of Otago, Dunedin, New Zealand\\
$^3$Department of Physics and Materials Science, Faculty of Science, Chiang Mai University, Chiang Mai 50200, Thailand
}%

\date{\today}

\begin{abstract}



Dysprosium (Dy) is the most magnetic element on the periodic table, making it excellent for studying dipolar atom-atom interactions. We report on a simple Dy MOT that captures atoms directly from the thermal beam using a single diode-laser system to generate the light. Additionally, the atoms are magnetically confined by the quadrupole magnetic field that also facilitates the MOT. The MOT loading time is $\tau_\text{b} = 26~\text{ms}$. Atoms can decay to a dark state that is magnetically trapped. The time constant for loading into this magnetic trap is $\tau_\text{d} = 410~\text{ms}$. The total magnetically trapped population is $1.14\times10^{5}$ atoms, with $85\%$ residing in the dark states. The magnetically trapped atoms have a temperature of $28~\mu\text{K}$, significantly below the Doppler limit. This population fulfills the requirements for a range of future experiments.


\end{abstract}

\maketitle


\section{\label{sec:intro}Introduction}

The precise control and manipulation of few atoms using light is advancing \cite{PhysRevLett.126.083401}, enabling groundbreaking new technologies such as quantum computing \cite{neutralatoms}. It also allows fundamental physics experiments that can directly observe events, revealing properties often concealed by large ensemble experiments \cite{Reynolds_2020}. Interest in laser cooling rare-earth elements has been rising over the last $15$ years \cite{Hostetter2015} due to their complex internal structures and rich spectral properties \cite{Lu2011}. Of these, Dy is of particular interest due to its large magnetic dipole moment \cite{dys}, allowing regimes where the atom-atom interaction is dominated by the magnetic dipole interaction rather than the typical electrostatic interaction \cite{Bloch2023}. Experiments in this regime provide insight into many-body effects such as quantum droplets \cite{quantum_droplets} and supersolids \cite{Tanzi2019}.

To date cold Dy experiments involve multi-stage laser cooling such as using 2-D magneto-optical traps (MOTs) \cite{Gao,Bloch2023} or a Zeeman slower \cite{levphd,Leo,Maier14,M_hlbauer_2018}. This typically requires complex setups with multiple lasers for different wavelengths that may involve second-harmonic generation of light \cite{levphd,Leo,Maier14,M_hlbauer_2018,Gao,Bloch2023}. The approaches are designed to produce large atom numbers and are likely excessive for a range of applications.

This work demonstrates a simple Dy MOT that only uses a single diode-laser system and no pre-cooling techniques. The MOT loads directly from a thermal beam of Dy atoms. Atoms in the MOT can decay to long-lived dark states in which they are decoupled from the cooling light. Atoms in dark states are confined by magnetic trapping facilitated by the quadrupole field also required for the MOT. We determine the time scales for the loading of the MOT and magnetic trap. The loading of atoms in bright states is $\tau_\text{b} = 26~\text{ms}$, and the lifetime of the dark states and hence also the time constant for loading is $\tau_\text{d} = 410~\text{ms}$.

The remainder of this paper is structured as follows: In Sec.~\ref{sec:Dysprosium}, we give the relevant properties of Dy and the cooling transition we use. Sec.~\ref{sec:over} gives an overview of our experiment setup used to form the MOT. We describe the $421~\text{nm}$ diode laser system in Sec.~\ref{sec:laser} before detailing our ultra-high vacuum (UHV) system in Sec.~\ref{sec:vacum}. In Sec.~\ref{sec:magfield} the magnetic field generation and fast switching is explained. Sec.~\ref{sec:MOT performance} shows that a Dy MOT is possible without pre-cooling, by estimating the proportion of atoms that will be trappable. Sec.~\ref{sec:count} describes how an on resonance beam is used to count the number of atoms in the MOT, before using this technique to optimize the MOT's controllable parameters in Sec.~\ref{sec:optimisation}. The rest of the results in Sec.~\ref{sec:results} show the loading rates for the MOT and magnetic trap and finds the total number of magnetically trapped atoms and their temperature. Lastly Sec.~\ref{sec:conc} concludes the paper, summarizing the main results.

\section{\label{sec:Dysprosium}Dysprosium's Properties}



Dysprosium has a roughly 50:50 mix of bosons to fermions, with $28.3\%$ being bosonic $^{164}$Dy \cite{Dy_isotopes} which is the isotope we focus on here. It has zero nuclear spin and no hyperfine structure omitting the need for re-pump lasers in laser cooling experiments \cite{LEVMOTPaper}. Dysprosium's ground state has J$=8$ with a g-factor of $1.24159$ and configuration of $4f^{10}(^5I_8)6s^2$. The $421~\text{nm}$ transition we use to cool Dy atoms with goes from the ground state to an excited state in $J=9$ ($4f^{10}(^5I_8)6s6p(^1P^o_1))$. The transition has a linewidth of $\gamma = 32.3~\text{MHz}$, a Doppler cooling limit of $T_{\text{Doppler}} = 774~\mu\text{K}$, and a saturation intensity of $I_{\text{sat}} = 564~\text{W}/\text{m}^2$ \cite{Lu2011}. Atoms in the excited state predominantly decay back to the ground state, but there is a possibility atoms can decay to other states, due to Dy's complex energy level diagram. Some of these states are long-lived so the atoms can spend an appreciable amount of time in dark states before they return to the ground state. 

Dysprosium has a magnetic dipole moment of $9.93$ $\mu_\text{B}$, the largest in the periodic table \cite{dys}. Due to Dy's low vapor pressure \cite{vapourpressure}, the sample needs to be heated to produce a significant atomic vapor.




\section{\label{sec:over}Overview of the Apparatus}

\begin{figure}[h]
	\centering	
	\includegraphics[width=0.49\textwidth]{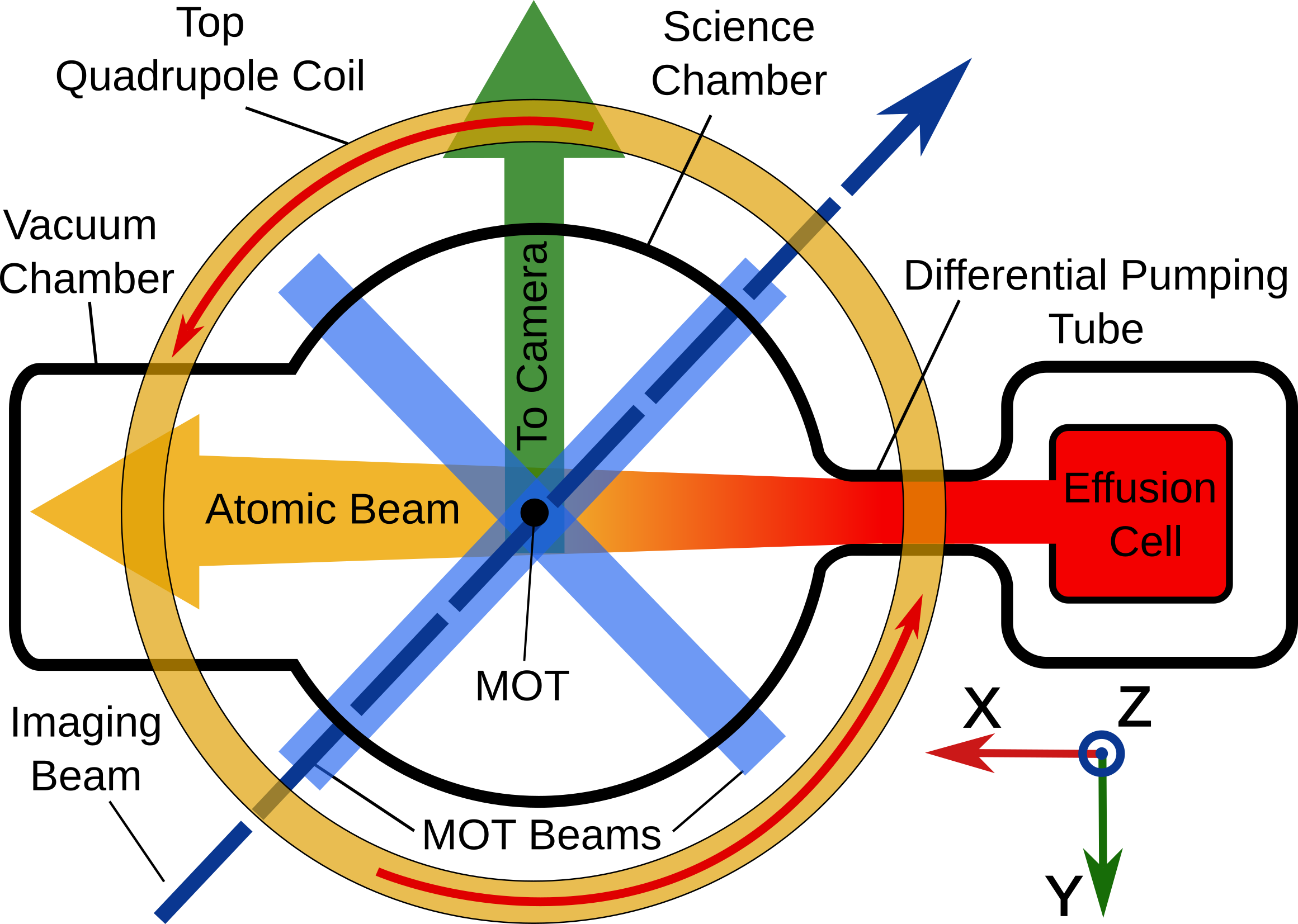}
	\caption{Overview of the Dy MOT and magnetic trap setup, showing how the effusion cell, vacuum chamber, quadrupole coils and three laser beams (the third is coming out of the page) are combined to form the Dy MOT. The quadrupole coil also provides the magnetic field for magnetic trapping.}
	\label{fig:overview}
\end{figure}

Fig.~\ref{fig:overview} shows a simplified schematic of the experiment where $10~\text{g}$ of Dy was placed into the effusion cell, shown on the right and typically heated to $1050~^\circ\text{C}$. The fast moving atoms are collimated by the differential pumping tube to form the hot atomic beam. The differential pumping tube isolates the science chamber where the MOT is formed from the effusion cell section of the UHV chamber. A pair of water-cooled quadrupole coils determine the center position of the MOT. These are directly  mounted to the outside of the vacuum chamber. The $3$ pairs of MOT beams; two of which are shown in Fig.~\ref{fig:overview} are $10~\text{mm}$ diameter and aligned to the center position of the chamber. When required, an "imaging beam" shown as a dark blue arrow along one of the MOT beam axis, induces the atoms to fluoresce. The fluorescence is imaged onto a low light sensitive camera \footnote{Xion Ultra 897 EMCCD} using an imaging system not shown in Fig.~\ref{fig:overview}. This setup allows us to form and image the Dy MOT, the details of which make up this paper.


\section{\label{sec:laser}421 $\text{nm}$ Laser System}



\begin{figure}[h] 
	\centering	
	\includegraphics[width=0.49\textwidth]{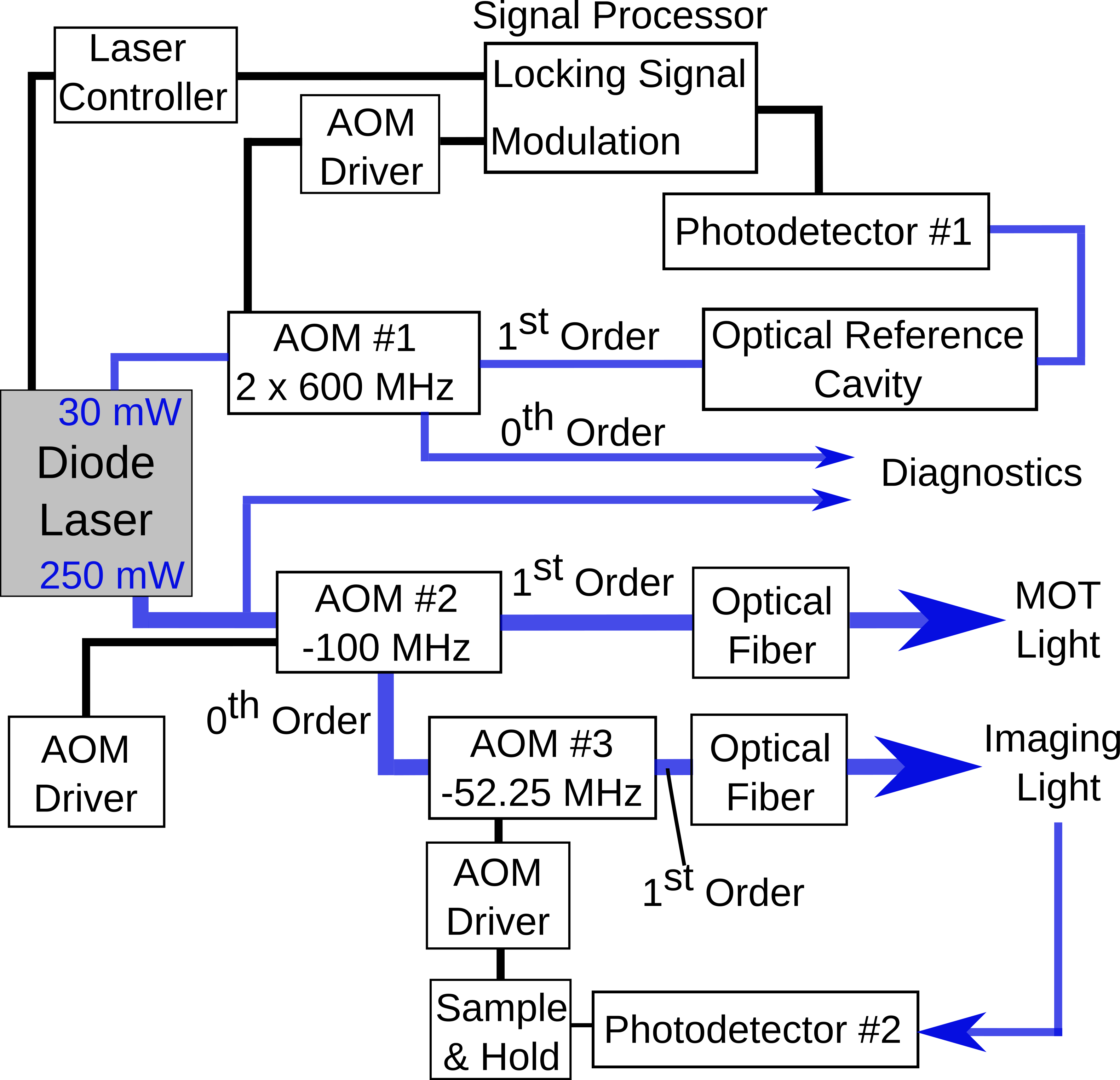}
	\caption{ Schematic of the laser system, used to form and analyze the Dy MOT, showing the distribution of the three laser beams. Also shown is the feedback loop for the cavity locking mechanism and imaging light power control}
	\label{fig:laseroverview}
\end{figure}

All light for our experiment is derived from a MOGLabs Injection-locked amplified diode laser system that outputs $421~\text{nm}$ light.  As indicated in the schematic in Fig.~\ref{fig:laseroverview} it outputs two beams: a low power ($40~\text{mW}$) beam and a high power output ($250~\text{mW}$). Following the beam path of the high power beam in Fig.~\ref{fig:laseroverview}, we first split a small amount of light off for diagnostics before downshifting the main beam using an acousto-optic modulator (AOM $\#2$) set at $100~\text{MHz}$. This downshifted beam becomes our MOT light. The zeroth order beam of AOM $\#2$ is then down shifted by $52.25~\text{MHz}$ by AOM $\#3$ to be at the Dy atomic resonance for imaging of the atoms.  We use both AOM $\#2$ and $\#3$ as fast shutters to switch the light on and off. Additionally AOM $\#3$ also stabilizes the imaging light power using a sample \& hold \cite{alma9924724663501891} based feedback loop.

To frequency stabilize the laser the $40~\text{mW}$ low power output in Fig.~\ref{fig:laseroverview} is double passed through a $600~\text{MHz}$ AOM before being coupled into a homemade optical reference cavity. Additionally the AOM frequency-modulates the light for the locking system. The cavity is $49.4~\text{mm}$ long with one plane mirror and one concave mirror with focal length $50~\text{mm}$. The mirrors are backside polished and have a reflectance of $99.83\%$ for $420~\text{nm}$ light, resulting in a measured linewidth of $2.5~\text{MHz}$, with resonant peaks spaced by $760\pm10~\text{MHz}$ \cite{hayton2024}. The cavity linewidth is hence significantly less than the atomic line width of $\gamma = 32.3~\text{MHz}$

The cavity temperature is actively stabilized using a thermoelectric based control system and isolated from the environment by a vacuum chamber pumped down to $5 \times 10^{-9}~\text{Torr}$. Photo-detector $\#1$, shown in Fig.~\ref{fig:laseroverview} detects the transmission signal which is converted into an error signal by a Teensy micro-controller based signal processor. The laser controller uses the error signal to correct the laser frequency as it drifts and keeps it locked to the cavity for days. While the reference cavity is made from a low thermal expansion glass it still has a small residual drift, meaning we need to correct the laser frequency for this every few hours.



\section{\label{sec:vacum}Ultra-High Vacuum Chamber}



\begin{figure}[h]
\centering	
\includegraphics[width=0.49\textwidth]{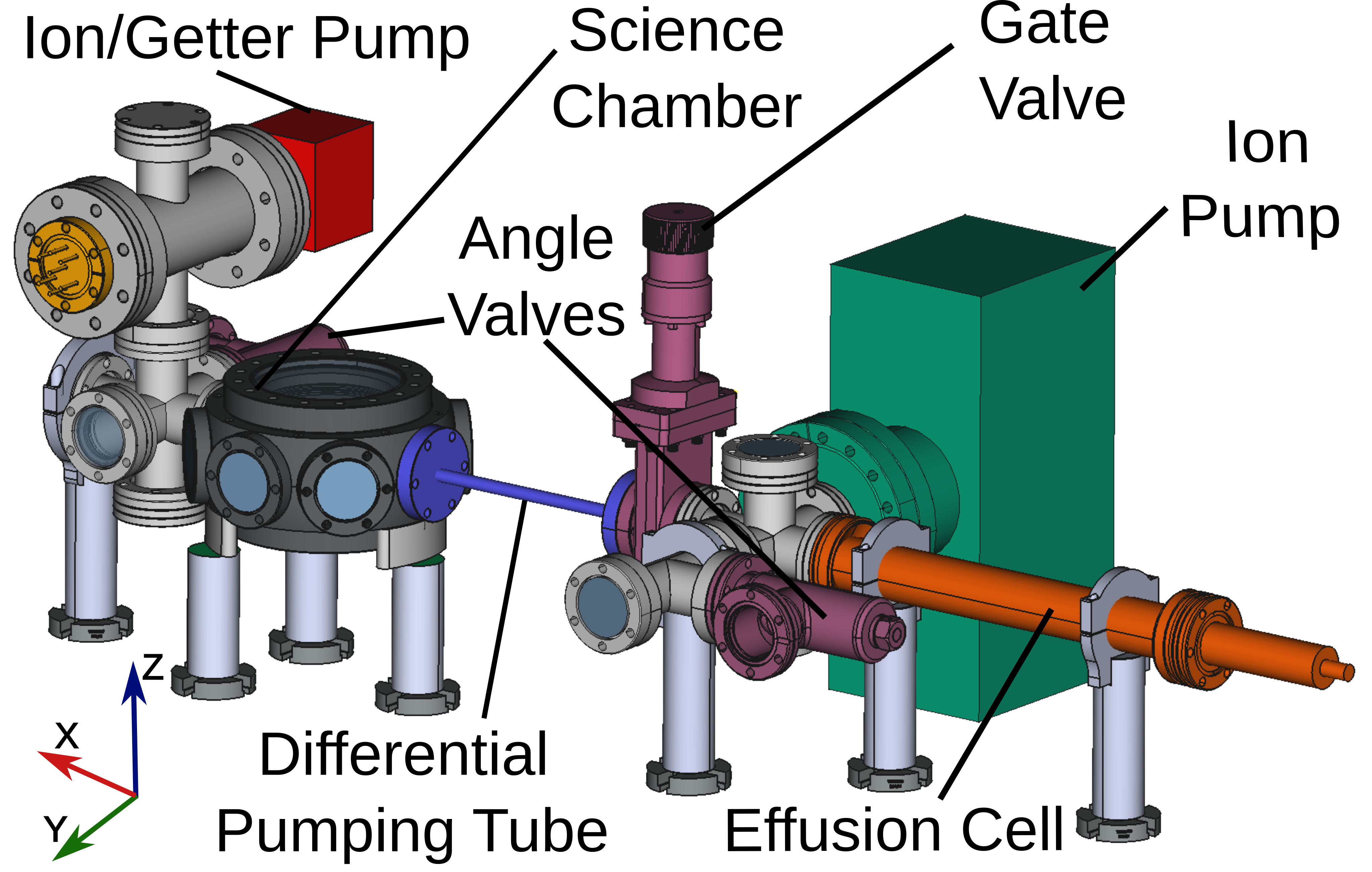}
\caption{CAD schematic of the vacuum chamber, showing the effusion cell in orange
	on the right, the science chamber in black on the left, separated by the differential pumping tube shown in blue}
\label{fig:vacchamber}
\end{figure}

Fig.~\ref{fig:vacchamber} depicts schematics of the vacuum chamber. It has two parts that are separated by the differential pumping-tube, colored in blue. The left part has the spherical-octagon science chamber, shown in black. It has eight $2.75"$ flanges around its circumference and two 6" flanges on the top and bottom. Connected to it is the red 500\text{L/s} ion/getter combination pump, which maintains the vacuum below $1.0 \times 10^{-11}~\text{mbar}$ when the effusion cell is off. Anti-reflective view-ports for $421~\text{nm}$ light are mounted to the science chamber. The science chamber and view ports are made from non-magnetic steel as the Dy atoms are sensitive to even small magnetic fields.

To the right of the differential pumping tube in Fig.~\ref{fig:vacchamber} is the effusion cell (shown in orange). It heats the Dy sample to cause atoms to sublimate. The green $75~\text{L/s}$ ion pump maintains the pressure at $1 \times 10^{-10}~\text{mbar}$, when the oven is heated to its working temperature of $1050~^\circ\text{C}$. The differential pumping tube is $20~\text{cm}$ long and has an internal diameter of $10~\text{mm}$ which allows for maintaining the pressure in the science chamber below $2 \times 10^{-11}~\text{mbar}$ while the effusion cell is running. A gate valve, shown in pink in Fig.~\ref{fig:vacchamber}, separates the two sides of the vacuum system allowing the effusion cell to be refilled without venting the science chamber. Each side has an angle valve attached also in pink, used for priming the vacuum.


\section{\label{sec:magfield}Magnetic Field Generation and switching}

\begin{figure}[h]
	\centering	
	\includegraphics[width=0.48\textwidth]{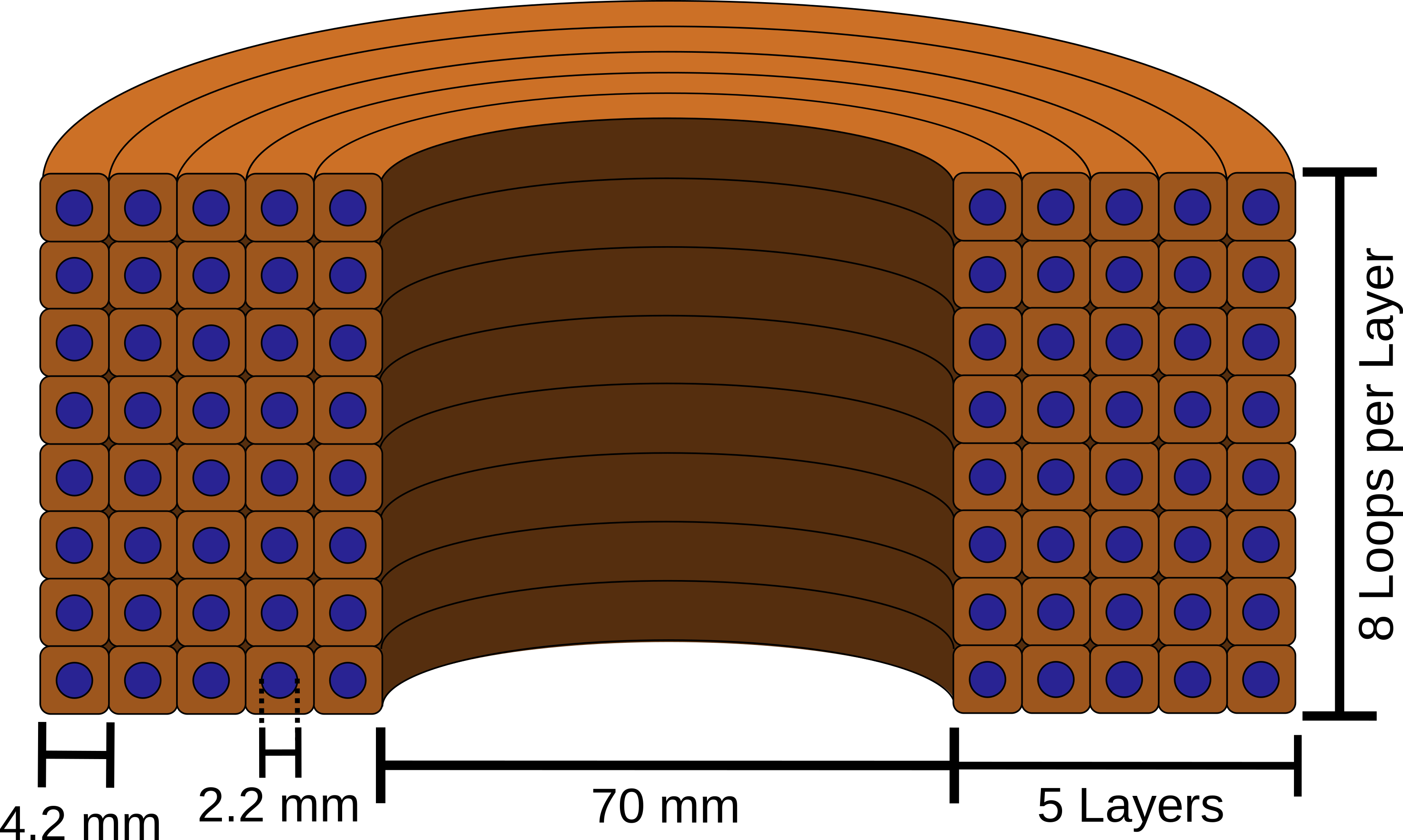}
	\caption{Cross-section of one quadrupole coil, showing the main dimensions and how the square hollow conductor is used for water cooling.}
	\label{fig:coilcros}
\end{figure}

To make a Dy MOT and a magnetic trap we need a quadrupole magnetic field, centered on the science chamber. Fig.~\ref{fig:coilcros} shows a cross-section of one of the water-cooled coils, showing each coil is made from square hollow conductor, with a width of $4.2~\text{mm}$. The water flows through the $2.2~\text{mm}$ center bore of the conductor at a rate of $1.3~\text{L/min}$ per coil to maintain the coils temperature during operation. To generate a magnetic field gradient of $36~\text{G/cm}$ along the z-axis, the coils are run at $130~\text{A}$, each coil having $5$ layers of $8$ loops and an internal diameter of $70~\text{mm}$. The coils are mounted on the top and bottom of the science chamber, and current in each coil is run in opposite directions to generate the quadrupole magnetic field.

\begin{figure}[h] 
\centering	
\includegraphics[width=0.48\textwidth]{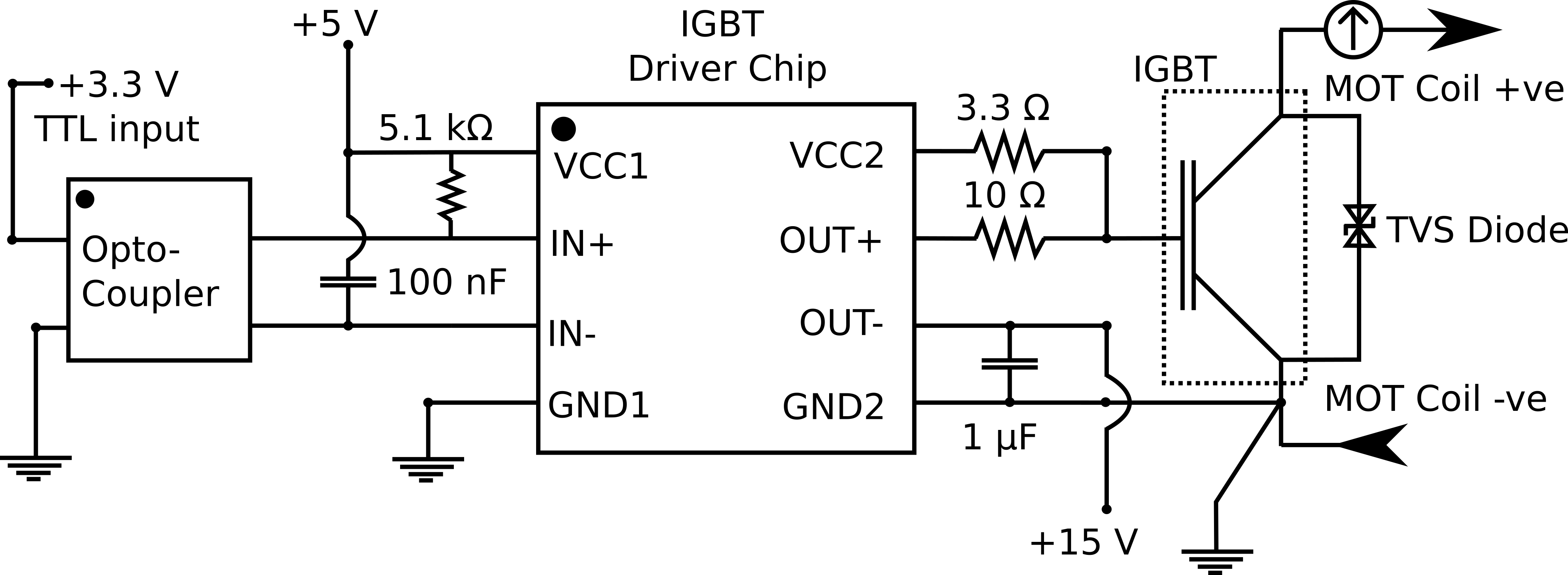}
\caption{Electrical circuit used to switch the quadrupole coils, showing the IGBT controller chip in the center and electrically isolating the circuit using an opto-coupler on the left. Also shown is the TVS diode on the right that prohibits voltage spikes.}
\label{fig:IGBTcircuit}
\end{figure}

Fast switching of the quadrupole field is essential for the optimization coming up in Sec.~\ref{sec:optimisation}. Fig.~\ref{fig:IGBTcircuit} shows how our insulated-gate bipolar transistor (IGBT) and its driver chip are combined following instructions from \cite{Infineon1}. On the right, connected to the IGBT and across the MOT coils, is a transient-voltage-suppression (TVS) diode that prohibits voltage spikes above $120~\text{V}$. With this circuit our coils can switch off from $150~\text{A}$ within $200~\mu\text{s}$. The Opto-coupler in Fig.~\ref{fig:IGBTcircuit} insulates the rest of the laboratory's electronics from potential high voltage transients.

Three pairs of smaller coils cancel static external magnetic fields in the science chamber by producing a uniform field on each axis. The $\hat{x}$ and $\hat{y}$ coils have $120$ loops of $1.08~\text{mm}$ wire, and an internal diameter of $73~\text{mm}$, whereas the z coils have $16$ loops of the same wire and a diameter of $183.2~\text{mm}$. All coil pairs are capable of producing a $2~\text{G}$ field in the center of the science chamber given at most $2~\text{A}$ of current. These coils are also mounted on the vacuum chamber.

\section{\label{sec:MOT performance}Atomic Source}

Our attention is now turned to the design considerations and characterizations of the atomic source. We start by calculating the capture velocity $v_{\text{cap}}$, the maximum speed an atom can have and still become trapped and use it to estimate the proportion of atoms from the source that travel slower than $v_{\text{cap}}$. Additionally the minimum speed $v_{\text{cut}}$ that the atoms leaving the effusion cell need to make it to the MOT region was determined.

\subsection{\label{sec:capv}Capture and Cut-off Velocities} 


The total force on an atom in a MOT in 1D is given by the sum of forces from two MOT beams,  $\mathbf{F}_{1D} = \mathbf{F}_{\mathbf{k}_1}+\mathbf{F}_{\mathbf{k}_2}$, the beam opposing the atom's motion ($\mathbf{k}_1$) and beam traveling in the same direction as the atom's($\mathbf{k}_2$), where \cite{Metcalf1999}
\begin{equation} 
	\mathbf{F}_{ \mathbf{k}}(x,v) =  \frac{\hbar \mathbf{k}s_0 \gamma/2}{1+s_0+(2(\delta -\mathbf{k} \cdot \mathbf{v} + \mu' A x/ \hbar )/\gamma)^2}.
	\label{eq:Fmot}
\end{equation}
In this equation $\mathbf{k}$ is the wave vector of the MOT beam, $s_0 = I/I_\text{sat}$ is the saturation parameter and $\gamma$ is the natural linewidth. In the denominator, $\delta = \omega_{\text{laser}}-\omega_{\text{atom}}$ is the detuning of the laser frequency from atomic resonance. $\mathbf{v}$ is the atom's velocity, $\mu'$ is Dy's effective magnetic moment, followed by the quadrupole field gradient $A$ and the atom's distance from the center of the trap $x$. The magnetic field gradient $A$ assumes the quadrupole field is zero in the center of the trap, where the MOT forms.


We numerically determine the capture velocity for a Dy atom. To do so we consider the atom's motion in 1D subject to a force from a pair of counter-propagating cooling beams. The force from each is given by Eq.~\ref{eq:Fmot}. The forces are present in a MOT region defined by our MOT beam diameter ($10~\text{mm}$). We assume the intensity to be constant, which is justified by the fact that the MOT beams are made from initial Gaussian beams with a $1/e^2$-diameter of $15~\text{mm}$ cropped by our optics to $10~\text{mm}$. For different initial velocities at the atoms entrance to the MOT region we compute the atoms trajectory using Euler's method. For high initial velocity the atom transitions through the MOT region without being stopped by the light forces. Decreasing the initial velocity until the light forces cause the atom to be trapped gives us the capture velocity. Since the atom's initial velocity is horizontal we ignore gravity in the calculation. We iteratively check the controllable parameters to predict favorable values. A good detuning was estimated to be $1.4 \gamma$ and an optimal magnetic field gradient to be $-29.5~\text{G/cm}$. The intensity of the light was also checked, but because the beam waist is set by our optics, the capture velocity scales directly with available laser power. With a maximum laser power available of $17~\text{mW}$ per beam and the other parameters optimized, we estimated our capture velocity to be $v_{\text{cap}} = 29~\text{m/s}$.

Another important velocity for the atoms is the minimum speed $v_\text{cut}$ they can travel and still enter the MOT region. Atoms traveling too slowly fall due to gravity and stick to the inside of the differential pumping tube or may fly below the MOT region. Considering this we estimate $v_\text{cut} = 5.5~\text{m/s}$ and atoms traveling slower than this cannot reach the MOT and can never be trapped.

\subsection{\label{sec:atomic source} Speed Distribution and Captured Fraction}

%



\begin{figure}[h] 
\centering	
\includegraphics[width=0.45\textwidth]{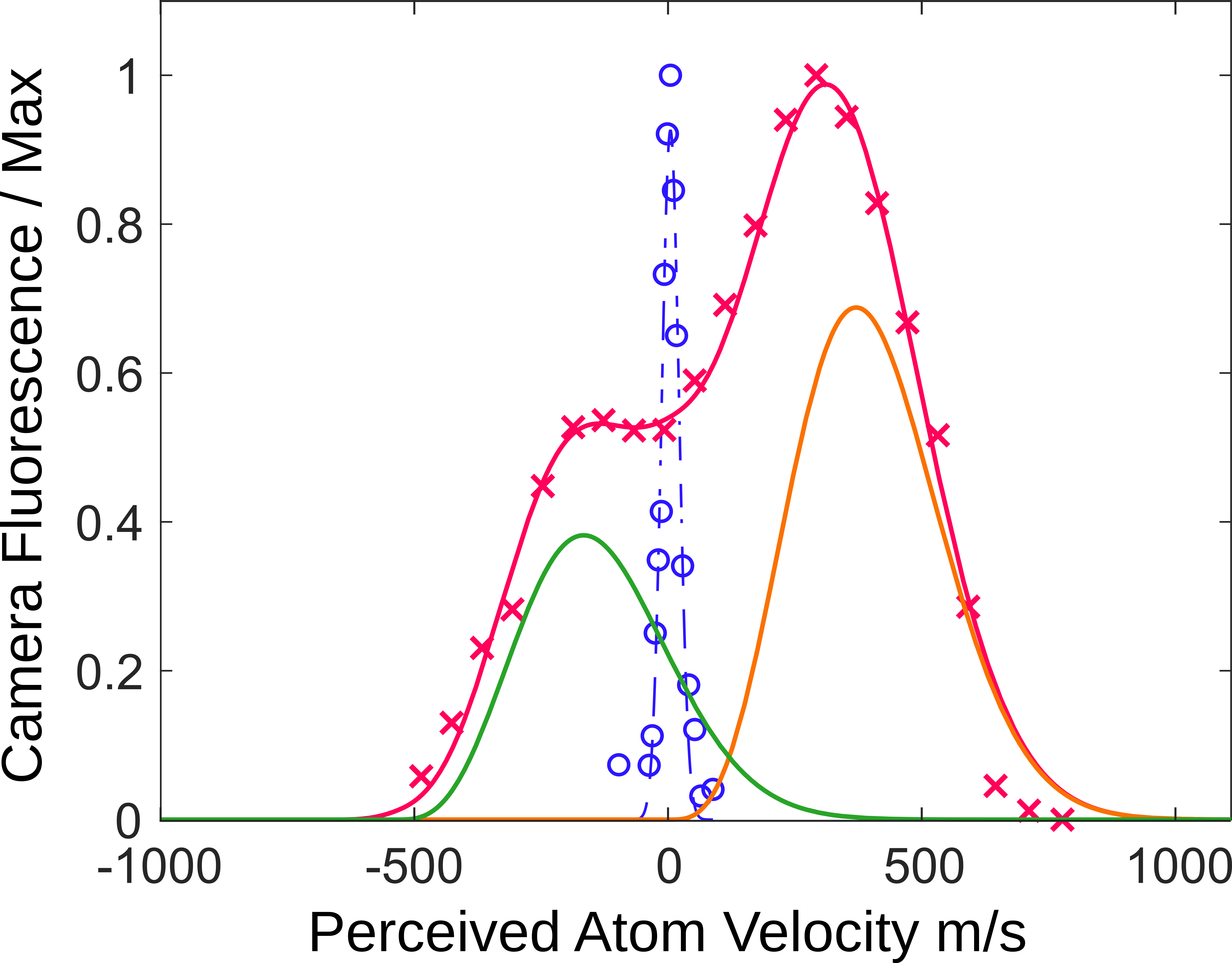}
\caption{Normalized fluorescence spectra showing the $^{164}$Dy Doppler free reference peak in purple and the Doppler broadened peak in red. We fit the red curve using a sum of distributions from $^{164}$Dy (orange), $^{162}$Dy (green) and $^{163}$Dy (not shown). The velocity axis is calibrated for the $^{164}$Dy peak only.}
\label{fig:atomspeed}
\end{figure}

To find the thermal beam's speed distribution, we measured the fluorescence of the atoms emerging in the science chamber using two different beams. The first beam is in the $\hat{z}$ direction, (see Fig.~\ref{fig:vacchamber}) and orthogonal to the atomic beam. Detecting fluorescence from the atoms induced by this beam yields the Doppler free transition frequency of $^{164}$Dy, shown as blue circles in Fig.~\ref{fig:atomspeed}. The second beam follows the same path as the imaging beam shown in Fig.~\ref{fig:overview}, pointing towards the atoms at a $45^\circ$ angle. The data collected with this beam is shown as red crosses in Fig.~\ref{fig:atomspeed} and shows the Doppler broadened distribution of multiple Dy isotopes. The laser frequency ($f_{\text{obs}}$) data is converted into perceived velocity ($v$) by using the previously determined transition frequency of $^{164}$Dy as our source frequency ($f_{\text{source}}$) in the Doppler frequency equation \cite{walker2007fundamentals}:
\begin{equation}
	v=\sqrt{2}c\frac{1-(f_{\text{obs}}/f_{\text{source}})^2}{1+(f_{\text{obs}}/f_{\text{source}})^2}.
	\label{eq:doppler}
\end{equation} 
The $\sqrt{2}$ in Eq.~\ref{eq:doppler} occurs as the laser beam makes a $45^{\circ}$ angle from the atomic beam, as shown in Fig.~\ref{fig:overview}.

Fitting the data shown in Fig.~\ref{fig:atomspeed} allows us to isolate the $^{164}$Dy velocity distribution and use it to estimate our capturable proportion. Each Dy isotope is assumed to have a velocity distribution of \cite{SchindlerMSc}:
\begin{equation}
	\mathcal{D}_{\alpha}(v)= A_{\alpha}\left(\frac{m_\alpha}{\sqrt{2}k_\text{B}T}\right)^2 v^3 \exp \left(-\frac{m_\alpha v^2}{2 k_\text{B} T}\right),
	\label{eq:modboltz}
\end{equation} 
where $A_{\alpha}$ is the relative abundance of an isotope ($\alpha$), $m_\alpha$ is the mass of the Dy atom, $T$ is the temperature of the atoms, $v$ is the atom's velocity. 

The fit to the data (red line) uses a sum of contributions from $^{162}$Dy, $^{164}$Dy and all of $^{163}$Dy's hyperfine states with velocity offsets ($\Delta_{\alpha}$) derived from the relative isotope shifts \cite{2009OptL...34.2548L},
\begin{equation}
f(v) = \mathcal{D}_{164}(v) + \mathcal{D}_{162}(v-\Delta_{162}) + \sum_{F} \mathcal{D}_{163,F}(v-\Delta_{163,F}),
\label{eq:fitfunchyper}
\end{equation}
where we have accounted for the hyperfine structure in $^{163}$Dy.
The green curve in Fig.~\ref{fig:atomspeed} shows the resulting speed distribution for $^{162}$Dy, and the orange is $^{164}$Dy. Integrating the $^{164}$Dy distribution from $v_{\text{cut}}$ to $v_{\text{cap}}$ gives us the probability of an atom in the MOT region being capturable to be $0.0045 \%$ or $1$ in every $22,000$ atoms.



\section{\label{sec:count}Determining the Atom Number}

We determine the atom number by comparing the known amount of light emitted by a Dy atom with the amount of light in a fluorescence image of the atomic cloud. The retro-reflected on-resonance imaging beam in Fig.~\ref{fig:overview} hits the MOT region and excites the atoms. The atoms decay back to the ground state and each emits light with an average power of $\hbar \omega_a \gamma/2$ as the imaging beam has $I \gg I_{\text{sat}}$. Of the emitted light, $2.87\%$ is collected within the solid angle set by a lens and imaged onto a low light sensitive camera. The calibrated camera allows us to record the amount of light that hits it during the exposure. With a known imaging pulse of $50~\mu\text{s}$ and correcting for the proportion of light collected by the imaging system we can then determine the atom number from an image. Determining the atom number using fluorescence only detects atoms in bright states.

\section{\label{sec:results}Experimental Results}


We start by identifying parameters that make the MOT achieve a good performance. This is followed by an investigation of the MOT loading dynamics that include how the populations of the bright and dark states evolve in time.

\subsection{\label{sec:optimisation}MOT Optimization}

 \begin{figure*}[t] 
	\includegraphics[width=0.8\textwidth]{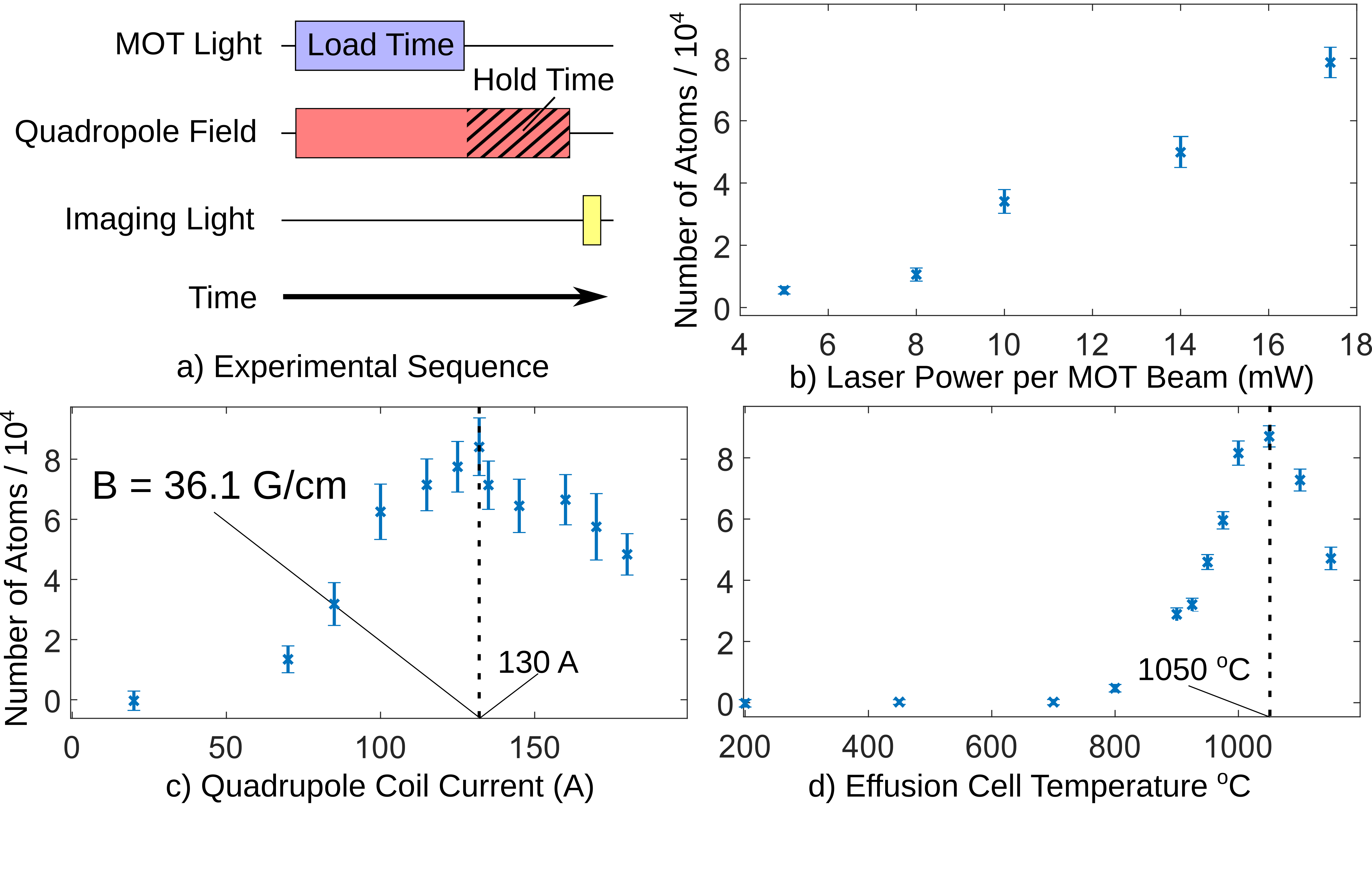}
	\caption{a) Showing the experimental sequence used to image the atoms with a load time of $2~\text{s}$ and a hold time of $1~\text{s}$. b), c) and d) shows optimization curves for our three main controllable parameters as follows; b) Laser power in the MOT beams, c) Quadrupole coil current and d) Effusion cell temperature as a function of atom number. Each point is the mean of $50$ background corrected images and the error bars denote the standard deviation in the sets of $50$}
	\label{fig:optimal}
\end{figure*}

 
We empirically found that a MOT light detuning of $49~\text{MHz}$ ($1.5\gamma$) gives our best result. This is consistent with the predicted optimal detuning in Sec.~\ref{sec:capv}. To further optimize parameters we run the experimental sequence illustrated in Fig.~\ref{fig:optimal}a, while varying the parameter under consideration. The sequence consists of loading the MOT for $2~\text{s}$ after which the light is shut off. The atoms are trapped in the quadrupole field for a wait time of $1~\text{s}$ after which the magnetic field is switched off within $200~\mu s$. Finally we determine the atom number. The wait time allows the magnetically trapped atoms in a dark state to return to the ground state and hence be detected. The optimal duration of the hold time ($t_{\text{hold}}=1~\text{s}$) is determined later on in Sec.~\ref{sec:MOTpop}.

Fig.~\ref{fig:optimal}b shows the atom number as a function of MOT beam power. The maximal atom number corresponds to our highest beam power available of $17.4~\text{mW}$. This indicates that a higher atom number could possibly be achieved if more laser power was available.
 
Fig.~\ref{fig:optimal}c shows how the atom number varies with quadrupole coil current. The maximal atom number corresponds to a coil current of $130~\text{A}$, which is equivalent to a gradient of magnitude $36.1~\text{G/cm}$ along the z-axis. Again the theoretically estimated parameter from Sec.~\ref{sec:capv} gave a good ballpark figure.
 
Lastly we varied the effusion cell temperature and present the results in Fig.~\ref{fig:optimal}d. A temperature of $1050~^\circ\text{C}$ yielded the highest measured atom number. Above this temperature the measured atom number drops. This could be due to the flux from the effusion cell becoming high which may result in significant atom loss during the $1~\text{s}$ hold time in the quadrupole magnetic trap.






\subsection{\label{sec:time scales}MOT Time-Scales}





\begin{figure}[h] 
	\includegraphics[width=0.45\textwidth]{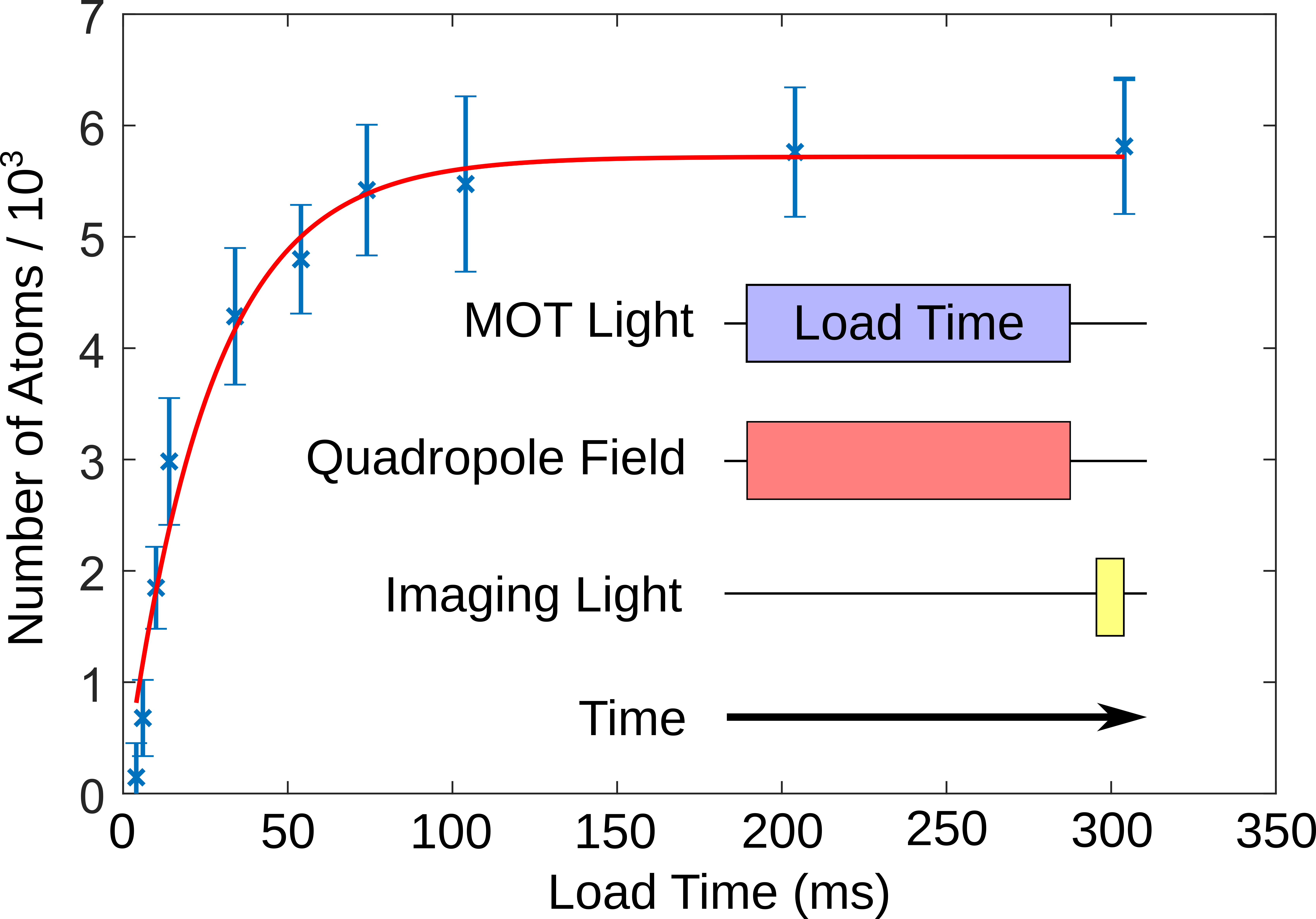}
	\caption{Bright state loading curve, showing the evolution of the bright states as we vary the MOT loading time. Each point is the mean of 50 background corrected images with 1 standard deviation error bars.}
	\label{fig:loading bright}
\end{figure}

The loading dynamics of the MOT is complicated by the previously mentioned dark states. This means that there are several time scales in play when atoms are loaded into the MOT. The first time scale is determined by the time it takes for the populations of atoms in the bright state to reach a steady state. We measured the time scale  using the experimental procedure outlined in the inset in Fig.~\ref{fig:loading bright}. It consists of loading the MOT for a variable duration immediately followed by determination of its atom number. Fig.~\ref{fig:loading bright} shows the experimental result. It is fitted with

\begin{equation}
	N = N_0 \left(1 - e^{-t/\tau}\right),
	\label{eq:MOTload}
\end{equation}
where $N_\text{o}$ is the saturated population, and $\tau$ is the time constant of the loading. Eq.~\ref{eq:MOTload} neglects atoms trapped in the dark states. The fit of Eq.~\ref{eq:MOTload} is the red solid line in the figure and it reveals the time constant of the bright state atoms to be $\tau_\text{b}  = 26~\text{ms}$ and $N_\text{o} = 5700$. The time constant is short due to atoms decaying into the dark states.


\begin{figure}[h]
	\centering	
	\includegraphics[width=0.45\textwidth]{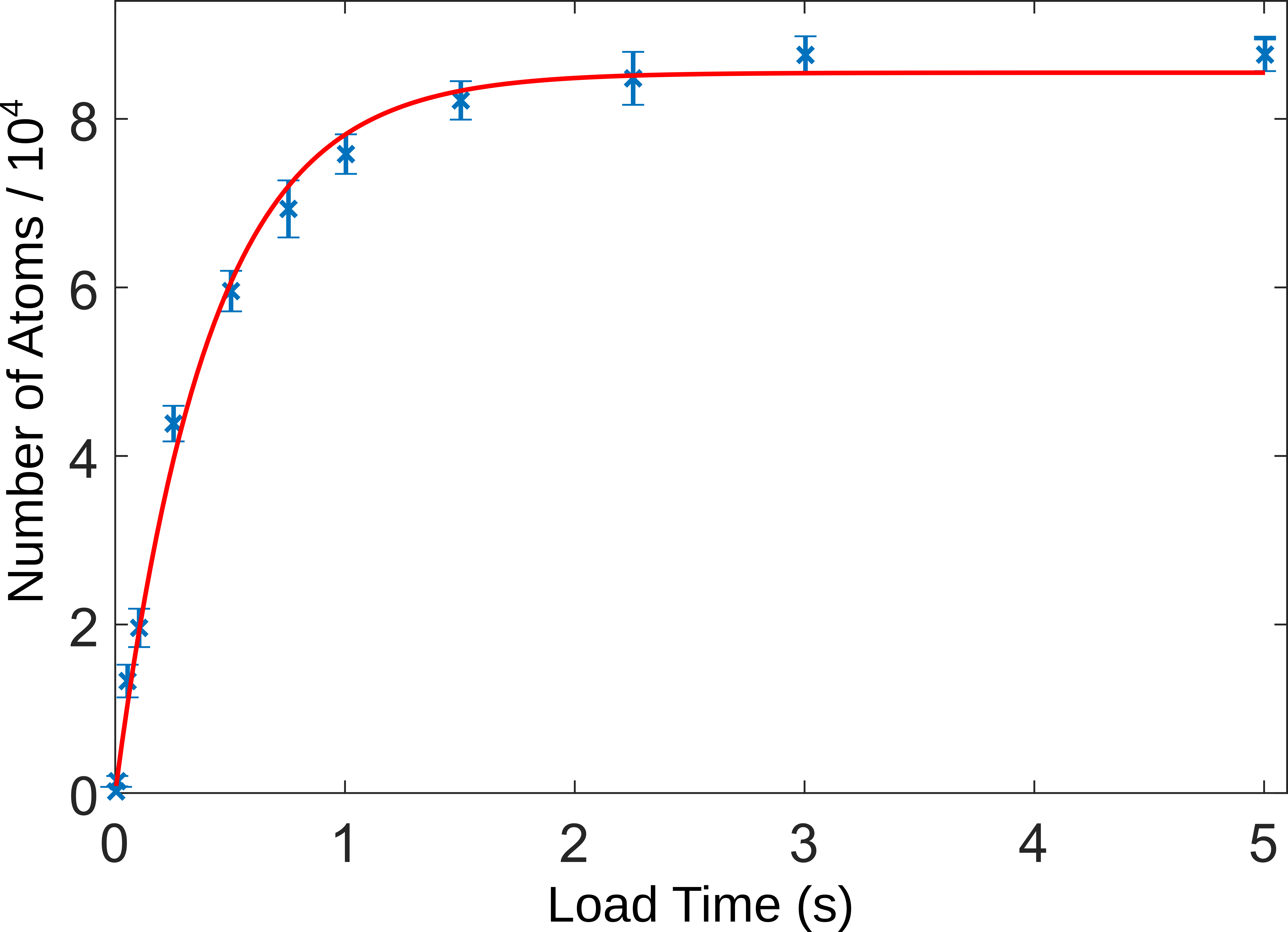}
	\caption{Dark state loading curve, showing the evolution of the dark states as we vary the MOT loading time, now with $t_{\text{hold}} = 1~\text{s}$. Each point is the mean of 50 background corrected images with one standard deviation error bars.}
	\label{fig:loading dark}
\end{figure}

Next we investigate the time scale for loading magnetically trapped dark state atoms. As illustrated in Fig.~\ref{fig:optimal}a, this is done by re-introducing the $1~\text{s}$ hold time in the quadrupole magnetic trap. This allows atoms in the dark state to decay back to the ground state where they can be detected. Fig.~\ref{fig:loading dark} shows the atom number measured in this way as a function of load time. The red line is a fit with Eq.~\ref{eq:MOTload}. The fit yields a $N_\text{o} = 85,000$ and comparing to Fig.~\ref{fig:loading bright} we see that approximately $90\%$ of atoms are magnetically trapped in the dark states. The fit reveals the time constant for loading atoms in the dark state is $\tau_\text{d} = 410~\text{ms}$ and the maximal observed number of atoms is reached with a load time of about $2~\text{s}$.

\subsection{\label{sec:MOTpop}MOT Population}

We have seen that as the MOT is loaded most atoms are magnetically trapped in the dark states. Some of these may decay to un-trapped sub-levels in the ground state and get lost. An interesting atom number is the total loaded number that can make it to trapped ground states. This number is denoted $N_\text{t}$ and it consists of the bright state atoms loaded into magnetically trapped ground states as well as the proportion of dark state atoms that will decay to the magnetically trapped ground state. 

\begin{figure}[h] 
	\centering	
	\includegraphics[width=0.30\textwidth]{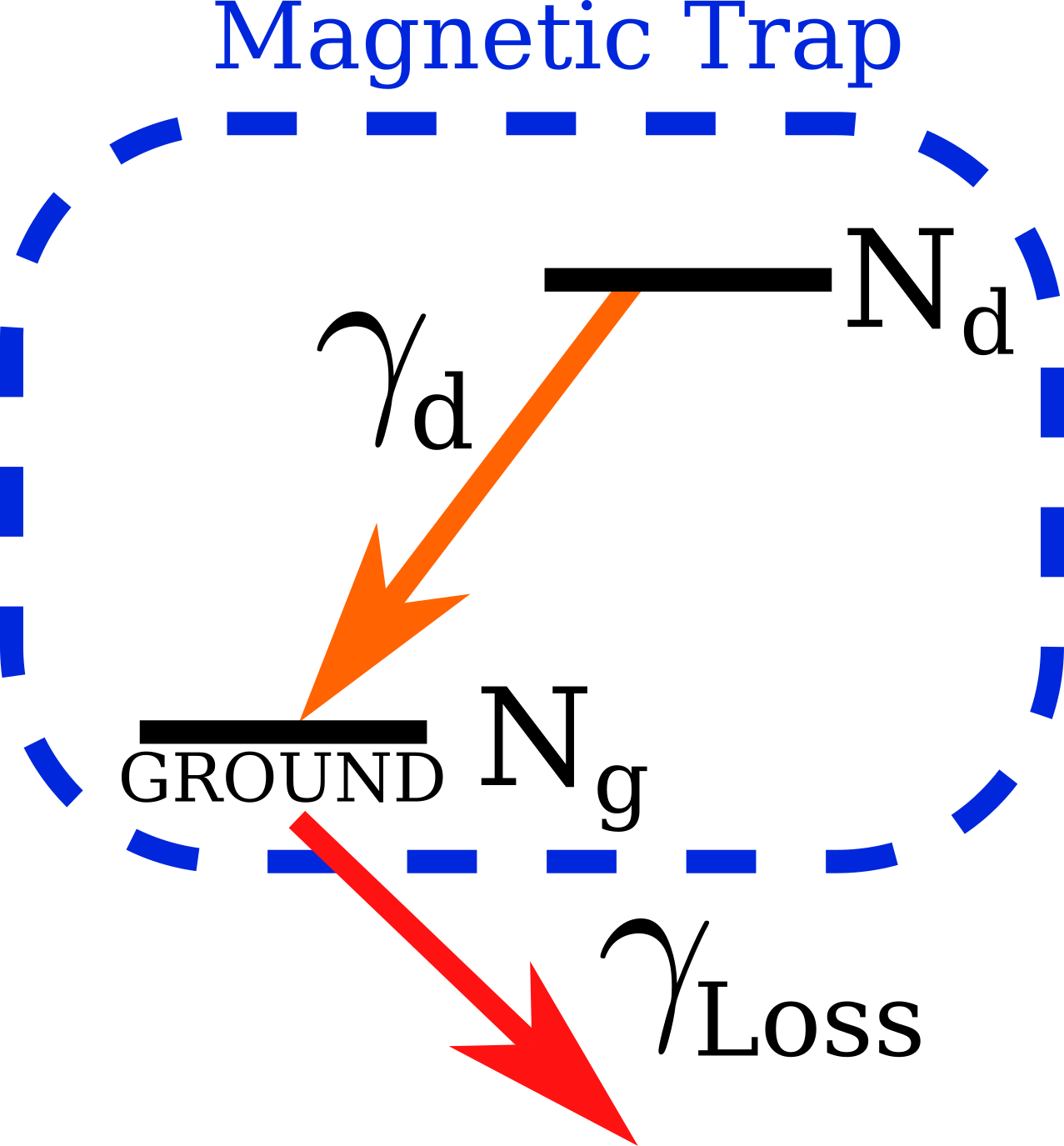}
	\caption{Two-level model that describes the atom's energy level dynamics while they are held in the quadrupole field only.}
	\label{fig:darkstates}
\end{figure}

\begin{figure}[h] 
	\centering	
	\includegraphics[width=0.49\textwidth]{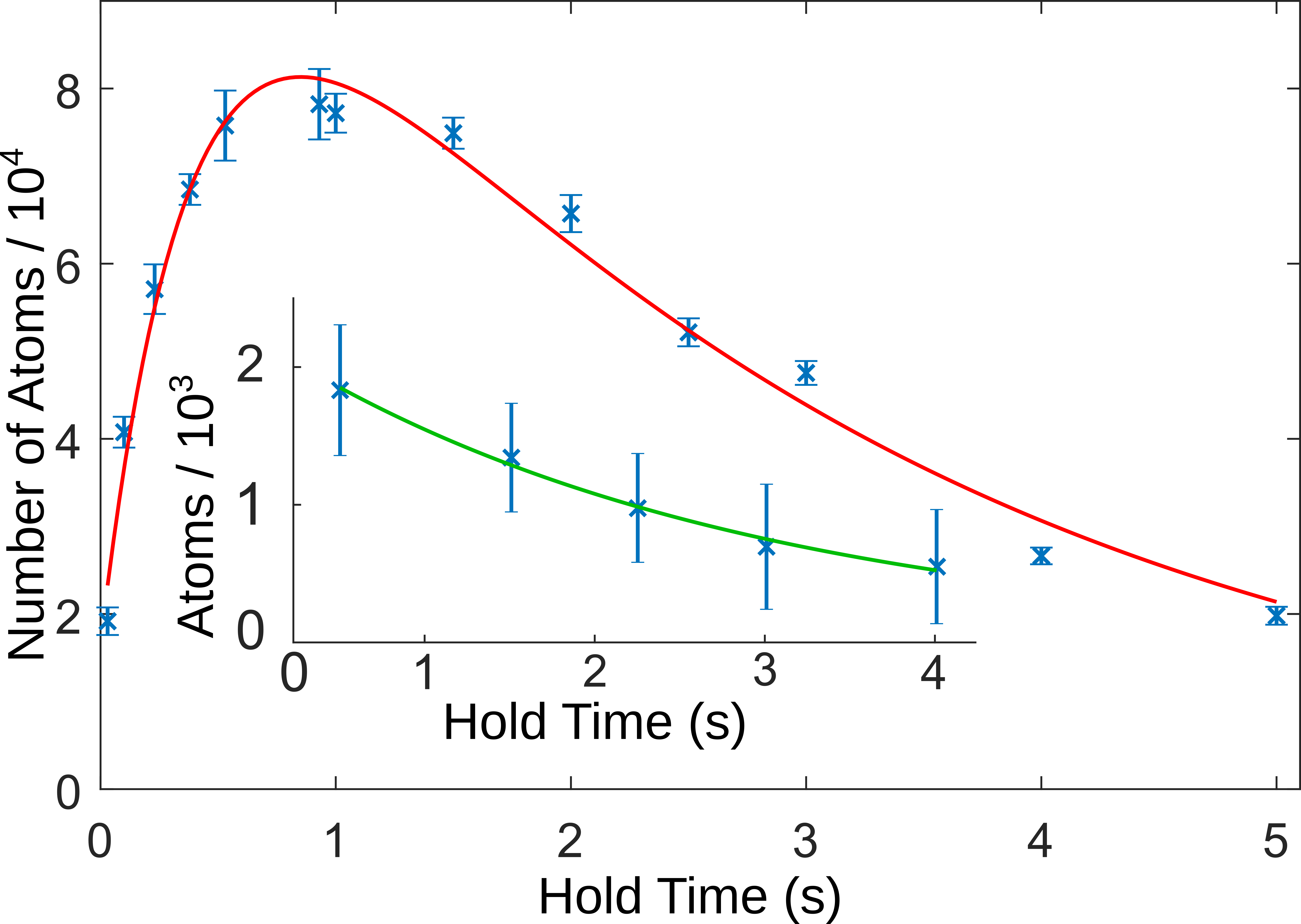}
	\caption{Evolution of atom number in the bright state as a function of hold time, with MOT loading time set at $t_{\text{load}} = 2~\text{s}$. Each point is the mean of 50 background corrected images with 1 standard deviation error bars.}
	\label{fig:hold time}
\end{figure}

To measure $N_\text{t}$ we consider the model illustrated in Fig.~\ref{fig:darkstates}. It contains two populations: $N_\text{g}$ which is the magnetically trapped atoms in the ground state and $N_\text{d}$ which is the population of dark state atoms that decay to magnetically trapped ground states. Additionally there is a decay rate $\gamma_\text{d}$ by which dark state atoms decay to the ground state and another decay rate $\gamma_\text{loss}$ by which magnetically trapped ground state atoms are lost, for example due to collisions with atoms in the thermal beam. This model leads to the following rate equations:
\begin{equation}
	\begin{aligned}
		\dot{N_\text{g}} &= -N_{\text{g}}\gamma_\text{loss} +N_\text{d} \gamma_\text{d}\\
		\dot{N_\text{d}} &= -N_\text{d} \gamma_\text{d}.
	\end{aligned}
\label{eq:rates}
\end{equation}
Solving these equations for $N_\text{g}(\text{t})$ gives:  
\begin{equation}
	N_\text{g}(t) = N_\text{g}(0)e^{-\gamma_\text{loss} t} + \frac{N_\text{d}(0)\gamma_\text{d}}{\gamma_\text{d}-\gamma_\text{loss}} (e^{-\gamma_\text{loss} t}-e^{-\gamma_\text{d} t}),
	\label{eq:N bright}%
\end{equation}
describing the populations of the bright state atoms as a function of time when MOT light is off. To determine $N_\text{t} = N_\text{g}+N_\text{d}$ we run the experimental sequence illustrated in Fig.~\ref{fig:optimal}a. Bright and dark states are loaded for $2~\text{s}$ before magnetically trapped atoms are held for a variable duration, after which we determine the atom number. The measured atom number peaks at a hold time duration of around $1~\text{s}$. The curve in the inset of Fig.~\ref{fig:hold time} shows data that determine the correct fit using a short load time to predominantly load atoms into the bright state, then measure their decay as they are held in the magnetic trap. The green line is a fit with $N_\text{i}e^{-\gamma_\text{loss}t_{\text{hold}}}$ where $N_\text{i} = 2,200~\text{atoms}$ is the number of atoms loaded in $10~\text{ms}$ and $\gamma_\text{loss} = 0.36\pm0.06~\text{s}^{-1}$. The error in $\gamma_{\text{loss}}$ is used as constraints for the fit of Eq.~\ref{eq:N bright} shown as red line in Fig.~\ref{fig:hold time} where $N_\text{g}(0)$, $N_\text{d}(0)$, $\gamma_\text{d}$ and $\gamma_\text{loss}$ (between $0.3~-~0.41~\text{s}^{-1}$) are fitting parameters. The fit revealed the results to be $N_\text{g}(0) = 17,000 $ atoms, $N_\text{d}(0) = 97,000$ atoms, $\gamma_\text{loss} = 0.36~\text{s}^{-1}$ and $\gamma_\text{d} = 2.4~\text{s}^{-1}$. The total number of magnetically trapped atoms is therefore $N_\text{t}(0) = 114,000$  and $85 \%$ of these are in the dark state. The highest number of magnetically trapped atoms in the ground state appear after $1~\text{s}$ hold time and is about $N_\text{g}(1~\text{s})  = 80,000$.




Lastly the temperature of the magnetically trapped atoms, measured after a $2~\text{s}$ hold time using a time-of-flight method \cite{timeofflight} is $T = 28~\mu\text{K}$. This is significantly below the Doppler temperature of $T_{\text{Doppler}} = 774~\mu\text{K}$. Sub-Doppler temperatures from a Dy MOT made using $421~\text{nm}$ light was previously observed in reference \cite{levtemp}.


\section{\label{sec:conc}Conclusion and Outlook}


In summary, we formed a MOT for Dy using only a single diode laser system and no pre-cooling techniques. The MOT captures $114,000$ atoms total, with a load time of $2$ seconds directly from the thermal beam. A large fraction ($85 \%$) of the atoms are magnetically trapped in meta-stable dark states. The temperature of the magnetically trapped atoms is $28~\mu\text{K}$, well below the Doppler limit. Although the atom number is small compared to setups that use pre-cooling it is sufficient for some experiments. In exchange for high atom number our setup is greatly simplified, making laser cooling of Dy more accessible.


\bibliography{bibliogrophy}

\end{document}